# Jeans and Boltzmann Solutions for Oblate Galaxies with Flat Rotation Curves

P.T. de Zeeuw[1], N.W. Evans[2] and M. Schwarzschild[3]

[1] *Sterrewacht Leiden, Postbus 9513, 2300 RA Leiden, The Netherlands*
[2] *Theoretical Physics, Department of Physics, 1 Keble Road, Oxford, OX1 3NP*
[3] *Princeton University Observatory, Peyton Hall, Princeton, NJ 08544, USA*

**ABSTRACT**

A general solution of the Jeans equations for oblate scale-free logarithmic potentials is given. This provides all possible second velocity moments that can hold up a stellar population of flattened scale-free density against the gravity field. A two-parameter subset of second moments for the self-consistent density of Binney's model is examined in detail. These solutions have the desirable property that the observable dispersions in the radial and proper motions can be given explicitly.

In the spherical limit, the potential of these models reduces to that of the singular isothermal sphere. The Jeans solutions for scale-free densities of arbitrary flattening that can correspond to physical three-integral distribution functions are identified.

The problem of finding distribution functions associated with the Jeans solutions in flattened scale-free logarithmic potentials is then investigated for Binney's model. An approximate solution of the collisionless Boltzmann equation is found which provides a third (partial) integral of good accuracy for thin and near-thin tube orbits. It is a modification of the total angular momentum. This enables the construction of many simple three-integral distribution functions. The kinematic properties of these approximate DFs are shown to agree with a subset of the Jeans solutions — which are thereby confirmed as good approximations to physical solutions. The observable properties of the models are discussed.

## 1 INTRODUCTION

Research in stellar dynamics applied to galaxies has generally followed one of two fairly distinct methods. In the first, the *statistical method*, phase space is divided into a six-dimensional grid of elementary cells and the stellar system is described by a six-dimensional distribution function (DF) for which the collisionless Boltzmann equation gives the continuity condition. In applying this method, the full DF is often replaced by its low-order moments, which are governed by the Jeans equations (e.g., Bacon 1985; Fillmore 1986; Evans & Lynden-Bell 1991; Magorrian & Binney 1994).

In the second, the *orbit method*, phase space is divided into a three-dimensional grid of three-dimensional slices, each of which contains one orbit. The stellar system is then described by a three-dimensional distribution function over all orbits (e.g., Schwarzschild 1979; Richstone 1980, 1984).

In the first method, the motion of a star is described only by its instantaneous location, velocity and acceleration. This is just enough to follow the star through an elementary cell, but not enough to determine the orbit family to which the star belongs, or whether the orbit might be stochastic. In contrast, the second method requires the determination of each orbit over its full extent in phase space, and thus provides a view of the structure of phase space for the potential under consideration.

The present paper applies exclusively the first, statistical, method. Within this method, we use approaches based on both the Jeans equations as well as the Boltzmann equation. The two approaches complement each other. The advantage of the Jeans equations is that they are directly related to observables, such as the shape and tilt of the velocity ellipsoid. But, the drawback is that such solutions are worthless unless they correspond to a physical (i.e., everywhere positive) DF satisfying the Boltzmann equation.

Numerical schemes are already under development (e.g., Dehnen & Gerhard 1993; Binney 1994) that may enable general DFs to be built for large classes of galaxy models. In this paper, our attention is concentrated on just one class - the oblate scale-free logarithmic potential (Richstone 1980; Binney 1981; Toomre 1982). This potential possesses the astrophysically important property of a flat rotation curve. The most widely-used member of the class — often called *Binney's model* — has spheroidal equipotentials and a simple self-consistent two-integral DF $f(E, L_z^2)$ (Toomre 1982; Evans 1993). We shall show here how to build simple approximate three-integral DFs as well. This supplements the earlier work of Richstone (1980, 1984) and Levison & Richstone (1985a, b), who found numerical DFs for this potential by the orbit method.

Section 2 describes the Jeans approach, and gives the complete solution of the Jeans equations for axisymmetric scale-free logarithmic potentials, as well as a useful two-parameter subset for the special case of Binney's model. Section 3 specialises these results to the limit of a spherical potential, and uses the Boltzmann approach to construct exact distribution functions. Section 4 describes the Boltzmann approach for flattened potentials, and presents a partial third integral which has fairly good accuracy in the part of phase space occupied by thin and near-thin tubes. The results from the two approaches are compared in Section 5. Finally, Section 6 describes how the observables can be obtained from the Jeans results.

## 2 THE JEANS APPROACH

We will work mainly in standard spherical coordinates $(r, \theta, \phi)$, with $\theta$ measured from the axis of symmetry and $\phi$ the azimuthal angle.

### 2.1 Jeans equations for spheroidal models

Consider an axisymmetric galaxy model with a gravitational potential $\Phi = \Phi(r, \theta)$ and with a phase-space DF $f = f(r, \theta, v_r, v_\theta, v_\phi)$. In equilibrium, $f$ must satisfy the collisionless Boltzmann equation (Binney & Tremaine 1987 [hereafter, BT], p. 279, eq. 4P-2):

$$0 = v_r \frac{\partial f}{\partial r} + \frac{v_\theta}{r} \frac{\partial f}{\partial \theta} + \left( \frac{v_\theta^2 + v_\phi^2}{r} - \frac{\partial \Phi}{\partial r} \right) \frac{\partial f}{\partial v_r}$$
$$+ \frac{1}{r} \left( v_\phi^2 \cot \theta - v_r v_\theta - \frac{\partial \Phi}{\partial \theta} \right) \frac{\partial f}{\partial v_\theta} \quad (2.1)$$
$$- \frac{v_\phi}{r} (v_r + v_\theta \cot \theta) \frac{\partial f}{\partial v_\phi}.$$

Multiplication by $v_r$, $v_\theta$, and $v_\phi$, respectively, and subsequent integration over velocity space, gives the Jeans equations. These relate the mass density associated with $f$

$$\rho = \iiint f \, dv_r \, dv_\theta \, dv_\phi, \quad (2.2)$$

and the elements of the stress tensor,

$$\rho \langle v_i v_j \rangle = \iiint f v_i v_j \, dv_r \, dv_\theta \, dv_\phi, \quad (2.3)$$

to the forces $\partial \Phi / \partial r$ and $r^{-1} \partial \Phi / \partial \theta$. In an axisymmetric model $\langle v_r v_\phi \rangle \equiv \langle v_\theta v_\phi \rangle \equiv 0$ by the symmetries of the individual orbits, and we are left with two non-trivial relations:

$$\frac{\partial \rho \langle v_r^2 \rangle}{\partial r} + \frac{1}{r} \frac{\partial \rho \langle v_r v_\theta \rangle}{\partial \theta}$$
$$+ \frac{\rho}{r} \left[ 2 \langle v_r^2 \rangle - \langle v_\theta^2 \rangle - \langle v_\phi^2 \rangle + \langle v_r v_\theta \rangle \cot \theta \right] = -\rho \frac{\partial \Phi}{\partial r}. \quad (2.4)$$
$$\frac{\partial \rho \langle v_r v_\theta \rangle}{\partial r} + \frac{1}{r} \frac{\partial \rho \langle v_\theta^2 \rangle}{\partial \theta}$$
$$+ \frac{\rho}{r} \left[ 3 \langle v_r v_\theta \rangle + (\langle v_\theta^2 \rangle - \langle v_\phi^2 \rangle) \cot \theta \right] = -\frac{\rho}{r} \frac{\partial \Phi}{\partial \theta}.$$

These two relations between the four stresses $\rho \langle v_r^2 \rangle$, $\rho \langle v_r v_\theta \rangle$, $\rho \langle v_\theta^2 \rangle$, $\rho \langle v_\phi^2 \rangle$ must be satisfied at any point in an axisymmetric equilibrium model with potential $\Phi$ and mass density $\rho$.

A solution of these equations for given $\rho$ and $\Phi$ corresponds to a physical equilibrium model only if it is associated with a DF $f(r, \theta, v_r, v_\theta, v_\phi) \geq 0$.

### 2.2 Logarithmic potentials and densities

Models with scale-free logarithmic potentials of the form

$$\Phi = \tfrac{1}{2} v_c^2 \ln[r^2 g(\theta)], \quad (2.5)$$

play a special role in modern astrophysics (e.g., Toomre 1982). This is because the circular velocity in the equatorial plane is constant and equal to $v_c$. For some applications, it is useful to consider tracer populations of stars moving in an external gravity field rather than the self-consistent stellar density implied through Poisson's equation. So, we take the density to have the general form:

$$\rho = \frac{h(\theta)}{r^\gamma g^2(\theta)}. \quad (2.6)$$

Here, $\gamma$ is a constant while $g(\theta)$ and $h(\theta)$ are arbitrary functions. Symmetry with respect to the equatorial plane requires $g(\pi - \theta) = g(\theta)$ and $h(\pi - \theta) = h(\theta)$. We could incorporate the factor $g^2(\theta)$ in the definition of $h(\theta)$, but it is convenient for what follows to write $\rho$ in the form (2.6).

When $\rho$ is the *self-consistent* density, then $\gamma = 2$ and $h(\theta)$ and $g(\theta)$ are related by

$$h(\theta) = g^2(\theta) + \tfrac{1}{2} g(\theta) g'(\theta) \cot \theta$$
$$+ \tfrac{1}{2} g(\theta) g''(\theta) - \tfrac{1}{2} g'(\theta) g'(\theta), \quad (2.7)$$

where we have taken $4\pi G = 1$ and $v_c = 1$. When $\gamma \neq 2$ the density (2.6) will overwhelm the density associated with the potential (2.5) either at large radii ($\gamma < 2$) or at small radii ($\gamma > 2$), so that results for this case should be used with caution.

The most widely-used logarithmic potential is the model studied by Toomre in the 1970s and subsequently popularised by Richstone (1980) and Binney (1981; see also BT, chap. 2). This has

$$g(\theta) = \sin^2 \theta + \frac{\cos^2 \theta}{q^2}, \quad (2.8)$$

so that the equipotentials are similar concentric spheroids with axis ratio $q$. The associated self-consistent density is of the form (2.6), with (BT, p. 46, eq. 2-54b)

$$h(\theta) = Q[2 - g(\theta)] = Q[1 + (1 - Q) \cos^2 \theta], \quad (2.9)$$

where $g(\theta)$ is given in (2.8), and we have written $Q = q^{-2}$. The density becomes negative when $Q > 2$. We shall refer to this potential and density as *Binney's model*.

Other cases of interest with potentials of the form (2.5) include the singular isothermal spheroids, which have $\rho = r^{-2}(\sin^2 \theta + Q \cos^2 \theta)$ and $g(\theta)$ is the integral given in equation (B2) of Qian et al. (1995). A family of scale-free models with elementary $g(\theta)$ and $h(\theta)$ and nearly spheroidal density distribution is given in Appendix C of de Zeeuw & Pfenniger (1988).

### 2.3 General solution of the Jeans equations

The potential (2.5) and density (2.6) have the desirable attribute of scale-freeness. Their properties at radius $r' = kr$ follow from those at radius $r$ by a simple magnification — for example, $\rho(kr,\theta) = k^{-\gamma}\rho(r,\theta)$. Here we consider DFs that are also scale-free, i.e., that satisfy $f(kr,\theta,v_r,v_\theta,v_\phi) = k^{-\gamma}f(r,\theta,v_r,v_\theta,v_\phi)$ (e.g., Richstone 1980; White 1985). The associated stresses therefore have the following form:

$$\rho\langle v_r^2\rangle = \frac{F_1(\theta)}{r^\gamma g^2(\theta)}, \qquad \rho\langle v_r v_\theta\rangle = \frac{F_2(\theta)}{r^\gamma g^2(\theta)},$$
$$\rho\langle v_\theta^2\rangle = \frac{F_3(\theta)}{r^\gamma g^2(\theta)}, \qquad \rho\langle v_\phi^2\rangle = \frac{F_4(\theta)}{r^\gamma g^2(\theta)}, \qquad (2.10)$$

where $F_1$, $F_2$, $F_3$, and $F_4$ are functions of $\theta$. They must fulfill the following conditions

$$F_1(\pi-\theta) = F_1(\theta), \qquad F_3(\pi-\theta) = F_3(\theta),$$
$$F_2(\pi-\theta) = -F_2(\theta), \qquad F_4(\pi-\theta) = F_4(\theta). \qquad (2.11)$$

These constraints guarantee that the stresses are symmetric with respect to the equatorial plane. That is, $\rho\langle v_r^2\rangle$, $\rho\langle v_\theta^2\rangle$ and $\rho\langle v_\phi^2\rangle$ must each have the *same* value at the point $(R,z) = (r,\theta)$ and its mirror image $(R,-z) = (r,\pi-\theta)$. But, $\rho\langle v_r v_\theta\rangle$ must have *opposite sign* at two such mirror points because it defines the tilt of the velocity ellipsoid with respect to the radial direction in a meridional plane $(R,z)$.

Substitution of the forms (2.10) into the Jeans equations (2.4) reduces them to two coupled first order ordinary differential equations:

$$F_2'(\theta) + \left[\cot\theta - \frac{2g'(\theta)}{g(\theta)}\right]F_2(\theta)$$
$$+ (2-\gamma)F_1(\theta) - F_3(\theta) - F_4(\theta) = -h(\theta), \qquad (2.12\text{a})$$

$$F_3'(\theta) + \left[\cot\theta - \frac{2g'(\theta)}{g(\theta)}\right]F_3(\theta)$$
$$+ (3-\gamma)F_2(\theta) - F_4(\theta)\cot\theta = -\frac{h(\theta)g'(\theta)}{2g(\theta)}. \qquad (2.12\text{b})$$

These equations provide two relations between the four functions $F_1(\theta),\ldots,F_4(\theta)$. We are free to pick two of the functions arbitrarily and solve (2.12) for the other two. We choose to prescribe $F_1(\theta)$ and $F_2(\theta)$, subject to the constraints (2.11) and the requirement $F_1(\theta) \geq 0$. This choice is motivated by the fact that it is then easy to consider the special case $F_2 = 0$, in which the velocity ellipsoid is everywhere aligned exactly along the spherical coordinate system.

We use equation (2.12a) to eliminate $F_4(\theta)$ from (2.12b). It is straightforward to integrate the resulting first order differential equation for $F_3$. Substitution of the result in (2.12a) then also gives $F_4$. We find:

$$F_3(\theta) = I(\theta) + J(\theta) + F_2(\theta)\cot\theta, \qquad (2.13\text{a})$$

$$F_4(\theta) = h(\theta) - I(\theta) - J(\theta)$$
$$- (\gamma-2)F_1(\theta) + g^2(\theta)\frac{d}{d\theta}\left[\frac{F_2(\theta)}{g^2(\theta)}\right], \qquad (2.13\text{b})$$

with

$$I(\theta) = \frac{g^2(\theta)}{\sin^2\theta}\int_0^\theta d\theta\frac{\sin^2\theta}{g^2(\theta)}\left[\cot\theta - \frac{g'(\theta)}{2g(\theta)}\right]h(\theta), \qquad (2.14)$$

and

$$J(\theta) = \frac{(\gamma-2)g^2(\theta)}{\sin^2\theta}\int_0^\theta d\theta\frac{\sin^2\theta}{g^2(\theta)}\left[F_2(\theta) - F_1(\theta)\cot\theta\right]. \qquad (2.15)$$

The stresses can hence be found by evaluation of the quadratures for $I(\theta)$ and $J(\theta)$. It follows that

$$F_3(0) = F_4(0) = \tfrac{1}{2}h(0) - \tfrac{1}{2}(\gamma-2)F_1(0), \qquad (2.16)$$

so that $\langle v_\theta^2\rangle = \langle v_\phi^2\rangle$ on the minor axis. Of course, this is required by axisymmetry — on the minor axis the $\phi$-direction is identical to the $\theta$-direction. This completes the derivation of the general solution of the Jeans equations for models with scale-free logarithmic potentials (2.5) and power-law densities of the form (2.6). It is particularly fortunate that the general solution can be written in simple analytical form.

Inspection of equations (2.13) shows that $I(\theta)$ and $J(\theta)$ occur only in the combination $I(\theta) + J(\theta)$, so that we could have combined them. However, the separate definitions of $I(\theta)$ and $J(\theta)$ shows the basic mathematical difference between solutions with $\gamma = 2$ and those with $\gamma \neq 2$. In the former case $J(\theta)$ vanishes, so that no integration of the arbitrary functions $F_1(\theta)$ and $F_2(\theta)$ is required. As we will mainly be concerned with this case, we give the simpler form of the general $\gamma = 2$ Jeans solutions explicitly:

$$F_3(\theta) = I(\theta) + F_2(\theta)\cot\theta,$$
$$F_4(\theta) = h(\theta) - I(\theta) + g^2(\theta)\frac{d}{d\theta}\left[\frac{F_2(\theta)}{g^2(\theta)}\right], \qquad (2.17)$$

where $I(\theta)$ is again given in equation (2.14). In this case we have $F_3(0) = F_4(0) = I(0) = \tfrac{1}{2}h(0)$.

### 2.4 Orientation of the velocity ellipsoid

The velocity ellipsoids defined by the general solution (2.13) are all aligned with the $\phi$-direction. But, in a plane of constant $\phi$, they are misaligned from the $r$- and $\theta$-direction when $\langle v_r v_\theta\rangle \neq 0$, i.e., when $F_2(\theta) \neq 0$. We define velocity components $(v_\lambda, v_\nu)$ in the meridional plane by

$$\begin{pmatrix} v_\lambda \\ v_\nu \end{pmatrix} = \begin{pmatrix} \cos\Theta & \sin\Theta \\ -\sin\Theta & \cos\Theta \end{pmatrix}\begin{pmatrix} v_r \\ v_\theta \end{pmatrix}, \qquad (2.18)$$

where we choose $\Theta$ at each position in such a way that $\langle v_\lambda v_\nu\rangle = 0$. This is accomplished by taking

$$\tan 2\Theta = \frac{2\langle v_r v_\theta\rangle}{\langle v_r^2\rangle - \langle v_\theta^2\rangle} = \frac{2F_2}{F_1 - F_3}, \qquad (2.19)$$

so that $\Theta$ is a function of $\theta$ only, as expected in a scale-free model. The transformation (2.18) diagonalises the stress tensor. The principal components $\rho\langle v_\lambda^2\rangle$ and $\rho\langle v_\nu^2\rangle$ are given by

$$\rho\langle v_\lambda^2\rangle = \frac{F_-(\theta)}{r^\gamma g^2(\theta)}, \qquad \rho\langle v_\nu^2\rangle = \frac{F_+(\theta)}{r^\gamma g^2(\theta)}, \qquad (2.20)$$

with

$$F_\pm(\theta) = \tfrac{1}{2}(F_3 + F_1) \pm \tfrac{1}{2}\sqrt{(F_3 - F_1)^2 + 4F_2^2}. \qquad (2.21)$$

A necessary — but not sufficient — condition for the stresses (2.10) to correspond to a physical equilibrium model is that $F_1, F_3, F_4 \geq 0$, since they give the velocity average of the non-negative quantities $v_r^2$, $v_\theta^2$, and $v_\phi^2$. We must also require $F_- \geq 0$ and $F_+ \geq 0$, which means we must have

$$F_2^2 \leq F_1 F_3, \qquad (2.22)$$

for the entire range of $\theta$ between 0 and $\pi/2$.

### 2.5 The two-integral limit: $f = f(E, L_z^2)$

We consider briefly the special case corresponding to a two-integral DF $f = f(E, L_z^2)$. In this model, the stellar velocities have no preferred direction in the meridional plane. In other words, $\langle v_r^2 \rangle \equiv \langle v_\theta^2 \rangle$ and $\langle v_r v_\theta \rangle \equiv 0$, so that the velocity ellipse in the $(R, z)$-plane is a circle. The associated stresses follow from the general solution by taking

$$F_1(\theta) \equiv F_3(\theta), \qquad F_2(\theta) \equiv 0. \qquad (2.23)$$

Substitution of (2.23) in equations (2.13a) and (2.15) results in an integral equation for $F_1$ which is of Volterra type. It is easily solved to give

$$F_1(\theta) = \frac{g^2(\theta)}{\sin^\gamma \theta} \int_0^\theta d\theta \, \frac{\sin^\gamma \theta}{g^2(\theta)} \Big[\cot \theta - \frac{g'(\theta)}{2g(\theta)}\Big] h(\theta). \qquad (2.24)$$

Substitution of this expression in equation (2.13b) then gives $F_4(\theta)$ as:

$$F_4(\theta) = h(\theta) + (1 - \gamma) F_1(\theta). \qquad (2.25)$$

When $\gamma = 2$, $F_1(\theta) = F_3(\theta) = I(\theta)$, and $F_4(\theta) = h(\theta) - I(\theta)$. These results can also be obtained from Hunter's (1977) solution of the Jeans equations for general $f(E, L_z^2)$ models.

The DF of the scale-free two-integral models is of the form $f(E, L_z^2) \propto \exp(-\gamma E) f_\gamma(\eta^2)$, where $\eta = L_z/L_c(E)$, and $L_c(E)$ is the value of $L_z$ on the circular orbit of energy $E$. The function $f_\gamma$ can be evaluated by means of the contour method of Hunter & Qian (1993). A number of specific cases can be found in the literature. Toomre (1982) and Evans (1993) give the elementary result for the self-consistent Binney model. Qian et al. (1995) give $f(E, L_z^2)$ for scale-free spheroidal densities in Binney's potential, and also for the self-consistent scale-free spheroids. They show that the second moments for the latter models are elementary when $\gamma = 2$.

### 2.6 A two-parameter solution for Binney's model

For the specific case of Binney's model, the entire general solution of the Jeans equations can be written down explicitly. We restrict ourselves to oblate models with $\frac{1}{2}\sqrt{2} \leq q \leq 1$ (i.e., $1 \leq Q \leq 2$.). The self-consistent density has $\gamma = 2$ so that $J(\theta) \equiv 0$. Taking $g(\theta)$ and $h(\theta)$ as given in (2.8) and (2.9) respectively, the integral (2.14) is elementary. We find

$$I(\theta) = \tfrac{1}{2}[\sin^2 \theta + Q(2-Q) \cos^2 \theta],$$
$$h(\theta) - I(\theta) = \tfrac{1}{2}[(2Q-1)\sin^2 \theta + Q(2-Q) \cos^2 \theta], \qquad (2.26)$$

Substitution in equation (2.17) then gives $F_3(\theta)$ and $F_4(\theta)$, for each choice of $F_2(\theta)$. Note that, when $\gamma = 2$, $F_1(\theta)$ does not even appear in the scale-free Jeans equations (2.12) and so its choice does not affect the other stresses.

Picking (2.23) for $F_1(\theta)$ and $F_2(\theta)$, and taking $\gamma = 2$, we recover the stresses for the two-integral $f(E, L_z^2)$ model derived by Evans (1993). They have a very simple functional dependence when written in terms of cylindrical coordinates $(R, \phi, z)$, with $R^2 + z^2 = r^2$ and $z = r \cos \theta$, namely

$$\rho \langle v_j^2 \rangle = \frac{a_j R^2 + b_j Rz + c_j z^2}{(R^2 + z^2/q^2)^2}, \qquad (2.27)$$

where $a_j, b_j$ and $c_j$ are constants which depend on $q$ ($j = R, \phi, z$). Evans & de Zeeuw (1994) showed that this simple form (2.27) has a desirable property — it allows an explicit evaluation of the line-of-sight projected second moments, such as the dispersions in the radial velocities and the proper motions. We elect to study the subset of the general solution (2.17) and (2.26) of the Jeans equations that possesses this pleasing characteristic. We now show that this defines a two-parameter family of anisotropic stresses for Binney's potential, and so this property is not restricted to the unique $f(E, L_z^2)$ model.

**Table 1.** Coefficients for the Jeans solution (2.30), with $Q = q^{-2}$.

| | |
|---|---|
| $A_1$ | $H_1 + H_2 + \tfrac{1}{2}$ |
| $C_1$ | $QH_2 + \tfrac{1}{2}Q(2-Q)$ |
| $B_2$ | $H_1$ |
| $A_3$ | $\tfrac{1}{2}$ |
| $C_3$ | $H_1 + \tfrac{1}{2}Q(2-Q)$ |
| $A_4$ | $Q - \tfrac{1}{2} - H_2$ |
| $C_4$ | $QH_2 + \tfrac{1}{2}Q(2-Q)$ |

Appendix A gives the transformation between the stresses in spherical and cylindrical coordinates. It is straightforward to show that all Jeans solutions (2.17) and

$$F_1(\theta) = I(\theta) + H_1 + H_2 \, g(\theta) \sin^2 \theta,$$
$$F_2(\theta) = H_2 \, g(\theta) \sin \theta \cos \theta, \qquad (2.28)$$

lead to stresses of the desired form (2.27). Here, $I(\theta)$ is given in equation (2.26) while $H_1$ and $H_2$ are constants. We emphasise that this is a very restricted subset of the general solution (2.17) — instead of two free *functions* $F_1(\theta)$ and $F_2(\theta)$, we have merely two free *parameters* $H_1$ and $H_2$. Solutions with $H_1 = 0$ correspond to velocity ellipsoids that are everywhere aligned with the cylindrical coordinate system. Solutions with $H_2 = 0$ are everywhere aligned with the spherical coordinate system.

Substitution of the choice (2.28) in the general solution (2.17) yields

$$F_3(\theta) = I(\theta) + H_2 \, g(\theta) \cos^2 \theta,$$
$$F_4(\theta) = h(\theta) - I(\theta) + H_2(Q \cos^2 \theta - \sin^2 \theta), \qquad (2.29)$$

with $I(\theta)$ and $h(\theta) - I(\theta)$ given in equation (2.26). It is helpful to write this Jeans solution out explicitly in cylindrical coordinates to emphasise its simplicity:

$$\rho \langle v_R^2 \rangle = \frac{A_1 R^2 + C_1 z^2}{(R^2 + z^2/q^2)^2},$$
$$\rho \langle v_R v_z \rangle = \frac{B_2 Rz}{(R^2 + z^2/q^2)^2},$$
$$\rho \langle v_z^2 \rangle = \frac{A_3 R^2 + C_3 z^2}{(R^2 + z^2/q^2)^2}, \qquad (2.30)$$
$$\rho \langle v_\phi^2 \rangle = \frac{A_4 R^2 + C_4 z^2}{(R^2 + z^2/q^2)^2},$$

where the relationships between the coefficients $A_j, B_j, C_j$ and our two anisotropy parameters $H_1$ and $H_2$ are given in Table 1.

Finally, to prepare for the comparison of the present results with those from the Boltzmann approach (Section 5.1), we assemble our solutions for Binney's model for the case $H_2 = 0$ in the following form

$$\rho \langle v_r^2 \rangle = \frac{1}{r^2 g^2(\theta)} \left[ \left[ \tfrac{1}{2}(2-Q) + H_1 q^2 \right] g(\theta) \right.$$
$$\left. + \sin^2 \theta (Q-1)(\tfrac{1}{2} + H_1 q^2) \right],$$
$$\rho \langle v_\theta^2 \rangle = \frac{1}{r^2 g^2(\theta)} \left[ \tfrac{1}{2}(2-Q) g(\theta) + \tfrac{1}{2}(Q-1) \sin^2 \theta \right], \quad (2.31)$$
$$\rho \langle v_\phi^2 \rangle = \frac{1}{r^2 g^2(\theta)} \left[ \tfrac{1}{2}(2-Q) g(\theta) + \tfrac{3}{2}(Q-1) \sin^2 \theta \right].$$

These formulae will be useful later.

### 2.7 Limits on the two-parameter solution

The principal stresses $\rho \langle v_\lambda^2 \rangle$ and $\rho \langle v_\nu^2 \rangle$ of the two-parameter Binney models described in the previous subsection are given in equation (2.20), where now

$$F_\pm(\theta) = I(\theta) + \tfrac{1}{2} H_1 + \tfrac{1}{2} H_2 g(\theta)$$
$$\pm \sqrt{H_1^2 - 2 H_1 H_2 g(\theta) \cos 2\theta + H_2^2 g^2(\theta)}. \quad (2.32)$$

The tilt angle $\Theta$ of the velocity ellipsoid in the meridional plane is given by

$$\tan 2\Theta = \frac{H_2 g(\theta) \sin 2\theta}{H_1 - H_2 g(\theta) \cos 2\theta}. \quad (2.33)$$

Necessary (but *not* sufficient) requirements for positive stresses are that $\rho \langle v_R^2 \rangle$, $\rho \langle v_z^2 \rangle$ and $\rho \langle v_\phi^2 \rangle$ are non-negative. This requires that $A_1 \geq 0, C_1 \geq 0, C_3 \geq 0$ and $A_4 \geq 0$, and translates into the following conditions on $H_1$ and $H_2$:

$$-\tfrac{1}{2} Q(2-Q) \leq H_1,$$
$$(\tfrac{1}{2} Q - 1) \leq H_2 \leq Q - \tfrac{1}{2}, \quad (2.34)$$
$$-\tfrac{1}{2} \leq H_1 + H_2.$$

These define a region in the $(H_1, H_2)$-plane illustrated in Fig. 1.

Necessary *and* sufficient conditions for positive stresses are that the principal components $\rho \langle v_\lambda^2 \rangle$ and $\rho \langle v_\nu^2 \rangle$ are non-negative. In fact, it is slightly easier to work with the sum and product of the principal components. The sum equals $\rho \langle v_R^2 \rangle + \rho \langle v_z^2 \rangle$, and hence is already positive, so only the product remains to be checked. This implies that at every point $(r, \theta, \phi)$, we need

$$\Delta(\theta) = a \sin^4 \theta + b \sin^2 \theta + c \geq 0, \quad (2.35)$$

where

$$a = (A_1 - C_1)(A_3 - C_3) + B_2^2,$$
$$b = C_1(A_3 - C_3) + C_3(A_1 - C_1) - B_2^2, \quad (2.36)$$
$$c = C_1 C_3.$$

The requirement (2.35) already holds at $\theta = 0$ and $\theta = \tfrac{1}{2}\pi$, since $D(0) = C_1 C_3$ and $D(\tfrac{1}{2}\pi) = A_1 A_3$. The Jeans solutions with negative stresses correspond to the cases when $\Delta(\theta)$ has a minimum in the range $0 < \theta < \tfrac{1}{2}\pi$ which is negative. Therefore, the additional regions that must be expunged from Fig. 1 are those simultaneously satisfying

$$a \geq 0, \quad 0 \leq -b \leq 2a, \quad -b^2 + 4ac \leq 0. \quad (2.37)$$

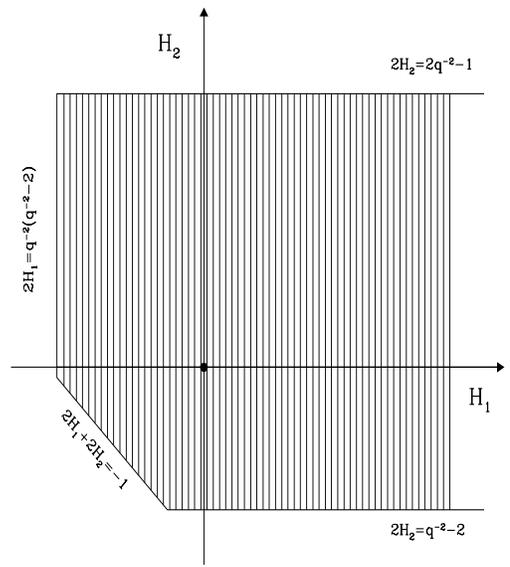

**Figure 1.** The $(H_1, H_2)$-plane of two-parameter anisotropic solutions of the Jeans equations for Binney's oblate scale-free logarithmic model with $\tfrac{1}{2}\sqrt{2} < q < 1$. The solutions with non-negative stresses $\rho \langle v_R^2 \rangle$, $\rho \langle v_z^2 \rangle$, and $\rho \langle v_\phi^2 \rangle$ lie in the hatched area. The requirement of positive principal stresses $\rho \langle v_\lambda^2 \rangle$ and $\rho \langle v_\nu^2 \rangle$ leads to the exclusion of two tiny regions near the two corners in the third quadrant. These are invisible on the scale of the Figure. The filled circle indicates the stresses with $H_1 = H_2 = 0$, which are associated with the $f(E, L_z)$ model. Spherically aligned velocity ellipsoids have $H_2 = 0$, while cylindrically aligned velocity ellipsoids have $H_1 = 0$. Approximate DFs for solutions with $H_2 = 0$ and $H_1 < 0$ are derived in Section 4.

Each of these four inequalities defines an area in the $(H_1, H_2)$-plane. The boundary curves follow by substituting the expressions in Table 1 in the definitions (2.36), and taking the equality signs in (2.37). This gives

$$H_2 = -\frac{(1-Q)^3}{2(1-Q)^2 - 4H_1},$$
$$H_2 = \frac{(1-Q)^2 \left[H_1 + Q(2-Q)\right]}{2(2Q-1)H_1 - Q(1-Q)(3-2Q)},$$
$$H_2 = \frac{(1-Q)^2 (H_1 + 1)}{2H_1 - (1-Q)(2-Q)}, \quad (2.38)$$
$$H_2 = (Q-1)H_1 \frac{\left[\sqrt{2H_1 + Q} \pm \sqrt{Q(2H_1 + 2Q - Q^2)}\right]^2}{\left[2H_1 + Q(1-Q)\right]^2}.$$

The area that lies inside all four boundaries needs to be excluded from the region defined by (2.34). Somewhat surprisingly, this results in the deletion of only two tiny areas near the two corners of the hatched region that lie in the third quadrant of Fig. 1. We will ignore them because they lie away from the Jeans solutions with $H_2 = 0$, for which we will construct approximate DFs in Section 4.

The position of the two-integral solution (2.23) is marked in Fig. 1 — it lies at the origin in the $(H_1, H_2)$-plane and corresponds to a positive definite DF (provided the flattening is not greater than $q = \tfrac{1}{2}\sqrt{2}$). As $H_1 \to \infty$, tended. It is clear from Fig. 1 that the set of Jeans solutions for Binney's self-consistent model is large. What is far from clear is which solutions are physically relevant.

## 3 SPHERICAL POTENTIALS

The existence of chaotic orbits prevents the extension of an analytic third integral to all the orbits in generic axisymmetric potentials. So, ascertaining which of the Jeans solutions have physical three-integral DFs is difficult in non-spherical models. We will consider this problem in Section 4, and here investigate the limiting case where the potential is exactly spherical. This allows the construction of galaxy models with three-integral DFs and triaxial kinematics. This is possible because the total angular momentum $L$ provides an exact third integral of motion in a spherical potential.

### 3.1 General Jeans solutions

In spherical symmetry, $g(\theta) \equiv 1$, so all the potentials (2.5) reduce to that of the singular isothermal sphere (BT, p. 228). The general solution for the stresses associated with a flattened density $\rho = r^{-\gamma} h(\theta)$ is of the form (2.10) with $g(\theta) \equiv 1$. The functions $F_1(\theta)$ and $F_2(\theta)$ are arbitrary, while $F_3(\theta)$ and $F_4(\theta)$ follow from (2.13) upon taking $g(\theta) \equiv 1$. However, it follows from the properties of the individual orbits in a spherical potential that $\langle v_r v_\theta \rangle \equiv 0$ for each of them. So, all solutions with $F_2(\theta) \not\equiv 0$ are unphysical! The remaining solutions have

$$F_3(\theta) = I(\theta) + J(\theta),$$
$$F_4(\theta) = h(\theta) - I(\theta) - J(\theta) - (\gamma - 2)F_1(\theta), \quad (3.1)$$

and

$$I(\theta) = \frac{1}{\sin^2 \theta} \int_0^\theta h(\theta) \sin\theta \cos\theta \, d\theta,$$
$$J(\theta) = \frac{(2-\gamma)}{\sin^2 \theta} \int_0^\theta F_1(\theta) \sin\theta \cos\theta \, d\theta, \quad (3.2)$$

where $F_1(\theta)$, which defines the angular variation of the radial velocity dispersion, is arbitrary. Once it and $h(\theta)$ have been chosen, the other stresses are fixed. In the special case $\gamma = 3$ we have $\langle v^2 \rangle \equiv \langle v_r^2 \rangle + \langle v_\theta^2 \rangle + \langle v_\phi^2 \rangle \equiv 1$.

Both $h(\theta)$ and $F_1(\theta)$ are even functions symmetric about $\theta = \pi/2$ and so each may be expanded as a Fourier series of the form $\sum a_{2\ell} \cos 2\ell\theta$ (see, e.g., Mathews & Walker 1964). These series can be rearranged as series in even powers of $\sin\theta$, so we may write without loss of generality:

$$h(\theta) = \sum_{n=0}^\infty a_n \sin^{2n} \theta, \qquad F_1(\theta) = \sum_{n=0}^\infty b_n \sin^{2n} \theta. \quad (3.3)$$

We now establish that it is possible to find DFs generating the Jeans solution (3.1) that corresponds to each pair $h(\theta), F_1(\theta)$. The specific case of scale-free spheroidal densities is discussed in some detail by de Bruijne, van der Marel & de Zeeuw (1996). Kochanek (1994) has also recently studied Jeans solutions of flattened denisties in the isothermal sphere.

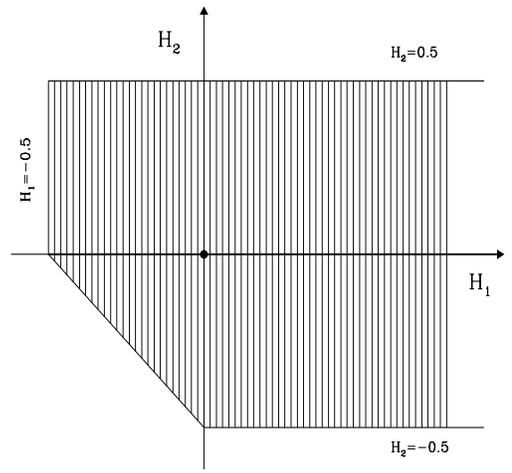

**Figure 2.** The $(H_1, H_2)$-plane of two-parameter anisotropic solutions of the Jeans equations for the singular isothermal sphere, which is the limit $q \to 1$ of Binney's model. The solutions with non-negative stresses lie in the hatched area. The filled circle indicates the stresses with $H_1 = H_2 = 0$, which are associated with the $f(E)$ isotropic model. The only physical solutions are those with spherically aligned velocity ellipsoids, which lie along the thick solid line $H_2 = 0$. The model built exclusively with circular orbits has $H_2 = 0$ and $H_1 = -1/2$, while the radial orbit model has $H_2 = 0$ and $H_1 \to \infty$.

White (1985) showed that the density $r^{-\gamma} \sin^{2n} \theta$ can be reproduced by non-negative DFs of the form

$$f_{m,n}(E, L^2, L_z^2) = \eta_{m,n} L^{2m} L_z^{2n} \exp[-(\gamma + 2n + 2m)E], \quad (3.4)$$

where $m+n > -1$ and $2n > -1$, and $\eta_{m,n}$ is a normalization constant given by

$$\eta_{m,n} = \frac{(m+n+\frac{\gamma}{2})^{m+n+\frac{3}{2}} \Gamma(n+1)}{\pi \Gamma(n+\frac{1}{2})\Gamma(n+m+1)}. \quad (3.5)$$

The corresponding stresses are

$$\rho \langle v_r^2 \rangle = \frac{1}{2m+2n+\gamma} \frac{\sin^{2n}\theta}{r^\gamma},$$
$$\rho \langle v_\theta^2 \rangle = \frac{m+n+1}{n+1} \rho \langle v_r^2 \rangle, \quad (3.6)$$
$$\rho \langle v_\phi^2 \rangle = (2n+1) \rho \langle v_\theta^2 \rangle.$$

For fixed $\gamma$ and $n$, the stresses and the density have the same angular dependence for all $m$, but the anisotropy ratios $\langle v_\theta^2 \rangle / \langle v_r^2 \rangle$ and $\langle v_\phi^2 \rangle / \langle v_r^2 \rangle$ do depend on $m$. White showed that further DFs can be constructed from (3.4) by linear superposition, and that the general solution of the Boltzmann equation is

$$f(E, L^2, L_z^2) = \sum_{m,n} \alpha_{m,n} f_{m,n}, \quad (3.7)$$

where the $\alpha_{m,n}$ are constants specifying the fraction contributed by each component. To recover the angular variation $h(\theta)$ of the density and $F_1(\theta)$ of the radial velocity dispersion, as given in equation (3.3), we therefore require

$$\sum_{m=0}^\infty \alpha_{m,n} = a_n, \qquad \sum_{m=0}^\infty \frac{\alpha_{m,n}}{2m+2n+\gamma} = b_n, \quad (3.8)$$

than constraints $a_n, b_n$. So, there are many ways to define three-integral DFs that reproduce the required velocity dispersions. Of course, we have not shown that these are all positive definite, although it is likely that at least some are – depending on the choice of $h(\theta)$ and $F_1(\theta)$. For example, de Bruijne et al. (1996) provide two families of three-integral DFs consistent with scale-free spheroidal densities of arbitrary flattening, which cover a large range in anisotropy, and are all physical.

### 3.2 A one-parameter solution for spherical models

It is instructive to consider in some detail the limiting behavior of the two-parameter Jeans solutions for the *self-consistent* Binney model discussed in Section 2.6. The density is now spherical, so that $h(\theta) \equiv 1$, and $I(\theta) \equiv \frac{1}{2}$. Since $\gamma = 2$ we have $J(\theta) \equiv 0$. We then find from equations (2.28) and (2.29):

$$
\begin{aligned}
F_1(\theta) &= \tfrac{1}{2} + H_1 + H_2 \sin^2\theta, \\
F_2(\theta) &= H_2 \sin\theta \cos\theta, \\
F_3(\theta) &= \tfrac{1}{2} + H_2 \cos^2\theta, \\
F_4(\theta) &= \tfrac{1}{2} + H_2(\cos^2\theta - \sin^2\theta).
\end{aligned} \tag{3.9}
$$

These Jeans solutions provide anisotropic stresses for the singular isothermal sphere. They are positive in the region in the $(H_1, H_2)$-plane defined by (cf. eq. [2.34])

$$
\begin{aligned}
&-\tfrac{1}{2} \leq H_1, \\
&-\tfrac{1}{2} \leq H_2 \leq \tfrac{1}{2}, \\
&-\tfrac{1}{2} \leq H_1 + H_2,
\end{aligned} \tag{3.10}
$$

which is illustrated in Fig. 2, and may be compared with Fig. 1. However, as we have seen, all physical solutions must have $F_2 \equiv 0$ because of the properties of individual orbits, so they all have $H_2 = 0$. We are left with a one-parameter solution

$$
\begin{aligned}
F_1(\theta) &= \tfrac{1}{2} + H_1, & F_2(\theta) &= 0, \\
F_3(\theta) &= \tfrac{1}{2}, & F_4(\theta) &= \tfrac{1}{2},
\end{aligned} \tag{3.11}
$$

which is indicated as the thick solid line in Fig. 2. It follows from the derivation in Section 3.1 that each of these Jeans solutions corresponds to a DF of the general form (3.7) with $n = 0$, so that (see also White 1985; Gerhard 1991; Kulessa & Lynden-Bell 1992)

$$ f(E, L^2) = \sum_{m=0}^{\infty} \alpha_{m,0} \eta_{m,0} L^{2m} \exp[-(2+2m)E], \tag{3.12} $$

with $\eta_{m,0}$ given in equation (3.5), and we must have

$$ \sum_{m=0}^{\infty} \alpha_{m,0} = 1, \qquad \sum_{m=0}^{\infty} \frac{\alpha_{m,0}}{m+1} = 1 + 2H_1. \tag{3.13} $$

Again, many DFs are possible. The simplest is obtained by taking only one component in the series (3.12):

$$ f(E, L^2) = \eta_{m,0} L^{2m} \exp[-(2+2m)E], \tag{3.14} $$

with

$$ m = -\frac{2H_1}{1+2H_1}. \tag{3.15} $$

This shows that $H_1$ is a measure of the anisotropy of the model. $H_1 = -\tfrac{1}{2}$ corresponds to the circular orbit model (no radial velocity dispersion) and $H_1 = 0$ to the isotropic model, which has $f = f(E)$. When $H_1 \to \infty$, the model approaches the radial orbit model. The DF (3.14) is non-negative and hence physical for each value of $H_1$ in the allowed range.

## 4 THE BOLTZMANN APPROACH

We now investigate which of the two-parameter Jeans solutions can correspond to three-integral DFs for the flattened Binney models.

### 4.1 Classical integrals

The collisionless Boltzmann equation (2.1) for the phase-space density $f$ can be put into the form

$$ 0 = v_r A + v_\theta B, \tag{4.1} $$

with

$$ A = \tfrac{1}{2} r \frac{\partial f}{\partial r} + (v_\theta^2 + v_\phi^2 - r\frac{\partial \Phi}{\partial r})\frac{\partial f}{\partial v_r^2} - v_\theta^2 \frac{\partial f}{\partial v_\theta^2} - v_\phi^2 \frac{\partial f}{\partial v_\phi^2}, \tag{4.2} $$

and

$$ B = \tfrac{1}{2}\frac{\partial f}{\partial \theta} - \frac{\partial \Phi}{\partial \theta}\frac{\partial f}{\partial v_\theta^2} + v_\phi^2 \cot\theta \left(\frac{\partial f}{\partial v_\theta^2} - \frac{\partial f}{\partial v_\phi^2}\right). \tag{4.3} $$

The derivatives of the potential (2.5) are (with $v_c = 1$)

$$ \frac{\partial \Phi}{\partial r} = \frac{1}{r}, \qquad \frac{\partial \Phi}{\partial \theta} = \frac{g'(\theta)}{2g(\theta)}. \tag{4.4} $$

This axisymmetric potential has two classical integrals, the energy $E = \Phi + \tfrac{1}{2}(v_r^2 + v_\theta^2 + v_\phi^2)$ and the $z$-component of the angular momentum $L_z = r v_\phi \sin\theta$. As is often done for convenience sake, we use these two integrals in the following form

$$
\begin{aligned}
I_1 &= e^{-2E} = \frac{1}{r^2 g(\theta)} e^{-(v_r^2 + v_\theta^2 + v_\phi^2)}, \\
I_2 &= L_z^2 = r^2 v_\phi^2 \sin^2\theta.
\end{aligned} \tag{4.5}
$$

These two integrals each fulfill the Boltzmann equation (4.1) by giving $A = 0$ and $B = 0$ separately.

### 4.2 A partial integral for Binney's model

We now restrict ourselves to the specific case of Binney's model. Its potential does not have an exact third integral, as is evident from the existence of some stochastic orbits (e.g., Richstone 1982). Here, no attempt is made to find an approximate third integral of reasonable accuracy for all orbits. Instead, our aim is to construct an approximate integral that has high accuracy for thin tube orbits. With such a *partial integral*, we can build models with tangentially anisotropic velocity distributions.

In the spherical limit ($q = 1$), the potential has an exact third integral, namely the total angular momentum $L^2 = r^2(v_\theta^2 + v_\phi^2)$. A number of authors (e.g., Saaf 1968; Innanen & Papp 1977) have suggested generalisations of $L^2$ as approximate third integrals for nearly spherical potentials. Here, we introduce a new modification, namely

$$ I_3 = L^2 g(\theta) = r^2(v_\theta^2 + v_\phi^2) g(\theta), \tag{4.6} $$

**Table 2.** Accuracy of the partial integral $I_3$. Meridional cross sections of the orbits are displayed in Fig. 3.

|     | $q$   | $R_0$  | $z_0$  | $L_z$ | $I_{3,\max}$ | $I_{3,\min}$ | $\delta I_3$ | $L^2_{\max}$ | $L^2_{\min}$ | $\delta L^2$ |
|-----|-------|--------|--------|-------|--------------|--------------|--------------|--------------|--------------|--------------|
|     | Thin tubes in E3 galaxies | | | | | | | | | |
| $A$   | 0.9   | .5507  | .2500  | .541  | .3678        | .3677        | .0001        | .3677        | .3535        | .0142        |
| $B$   | 0.9   | .4000  | .4500  | .378  | .3668        | .3655        | .0013        | .3655        | .3232        | .0423        |
| $C$   | 0.9   | .2000  | .5674  | .182  | .3649        | .3616        | .0033        | .3616        | .2992        | .0624        |
| $D$   | 0.9   | .0500  | .5997  | .045  | .3640        | .3599        | .0041        | .3599        | .2920        | .0679        |
|     | Thin tall tubes in flat galaxies | | | | | | | | | |
| $D^*$   | 0.8   | .0500  | .5838  | .040  | .3506        | .3333        | .0173        | .3333        | .2147        | .1186        |
| $D^{**}$ | 0.707 | .0500  | .5570  | .034  | .3282        | .2905        | .0377        | .2905        | .1479        | .1426        |
|     | Fat tubes in E3 galaxies | | | | | | | | | |
| $E$   | 0.9   | .4800  | .5000  | .378  | .3586        | .3318        | .0268        | .3580        | .2969        | .0611        |
| $F$   | 0.9   | .6600  | .4800  | .378  | .2754        | .2338        | .0416        | .2705        | .2187        | .0518        |
| $G$   | 0.9   | .8400  | .3000  | .378  | .1805        | .1634        | .0171        | .1781        | .1607        | .0174        |

with $g(\theta) = (\sin^2 \theta + Q \cos^2 \theta)$. Inserting this $I_3$ into the Boltzmann equation (4.1), we find that

$$A = 0, \qquad B = \tfrac{1}{2}r^2(v_\theta^2 + v_\phi^2 - 1)g'(\theta). \qquad (4.7)$$

As $g'(\theta)$ is non-zero for a flattened model, $I_3$ is an approximate integral for those orbits for which the condition

$$v_\theta^2 + v_\phi^2 \approx 1, \qquad (4.8)$$

is fulfilled.

The easiest way to investigate the accuracy of the apporximate integral is to numerically integrate representative orbits and find the fluctuation in $I_3$. The results of such computations are listed in Table 2, while Fig. 3 shows cross-sections of the orbits in the meridional plane. All the orbits are normalised to $E = 0$ and are identified by their starting values $R_0$ and $z_0$, with $v_r$ and $v_\theta$ zero there. In each case, the maximum and minimum values of $I_3$ along the orbit are recorded, together with the variation in $I_3$ and in $L^2$. The top section of this Table gives the data for four thin tubes in a mildly non-spherical potential ($q = 0.9$). For all these thin tubes — even for the last one with the highest $z$-excursion — the fluctuations in $I_3$ are small. The middle section of the Table contains two thin tubes with high $z$-excursions, but for strongly flattened potentials ($q = \tfrac{1}{2}\sqrt{2}$ is the limit at which the density on the $z$-axis vanishes). For the flattest potentials, the $I_3$ excursions are becoming uncomfortably large. Finally, in the bottom section of the Table, three fat tubes — one very fat — show similarly large excursions in $I_3$. The last column in Table 2 gives the variation in $L^2$ for the same nine orbits. Comparison of this column with the one for $\delta I_3$ shows that the excursions of $I_3$ are persistently smaller than those of $L^2$, in fact by a factor exceeding ten for all thin tubes with $q = 0.9$. We conclude that $I_3$ is indeed a *partial integral*, in the sense that it is quite accurately conserved on thin and near-thin tube orbits in Binney's model with $q \gtrsim 0.8$. It is not so well conserved for fat tubes.

Table 2 contains a sequence of four orbits (B, E, F, and G) which all have $E = 0$ and $L_z = 0.378$. They therefore can not be discriminated by the classical integrals, even though they have distinctly different shapes, as shown in Figure 3, running from exactly thin (B) to very fat (G). But the approximate integral $I_3$ can discriminate between them. Figure 4 shows the range in $I_3$ ($I_{3,\min}$ to $I_{3,\max}$ of Table 2) for each of the four orbits B, E, F, and G. These ranges do not overlap. This shows that the average value of $I_3$ for each orbit is a useful discriminant between orbits of the same $E$ and $L_z$. The bottom half of Figure 4 indicates that $L^2$ is a less effective discriminant than $I_3$, particularly between thin and near-thin orbits.

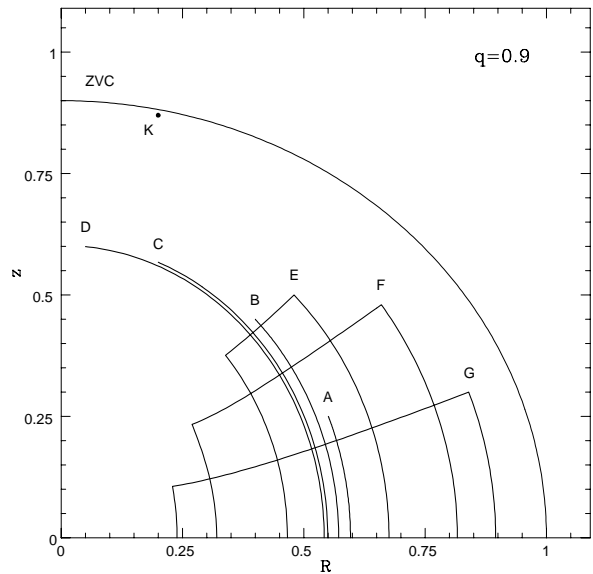

**Figure 3.** Cross sections of the orbits of Table 2 with a meridional plane $(R, z)$ for the potential of Binney's oblate scale-free model with $q = 0.9$ at energy $E = 0$. The zero-velocity curve (for $L_z = 0$) is indicated by ZVC. Orbits A – D are thin tubes. Orbits E – G are fat (or thick) tubes that fill the indicated areas. The dot indicates the upper turning point of orbit K, which fills most of the area inside the ZVC (see section 5.2).

### 4.3 Three-integral distribution functions

The two exact integrals $I_1$, $I_2$ and the partial integral $I_3$ can be used to construct approximate DFs $f(I_1, I_2, I_3)$. We define a basic solution of the Boltzmann equation to be a

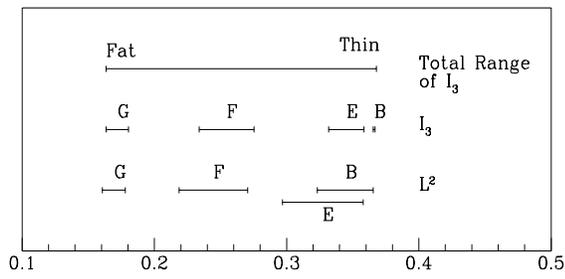

**Figure 4.** The variations of $L^2$ and $I_3$ for orbits B, E, F, and G, which all have the same values of $E$ and $L_z$. The average value of the approximate integral $I_3$ distinguishes all orbits quite clearly. $L^2$ is a much less useful discriminant. The exact range of $I_3$ and $L^2$ for each orbit is given in table 2.

product of powers of the three integrals (cf. White 1985; Section 3.1)

$$f_{k,n,m} = I_1^k I_2^n I_3^m = \frac{1}{r^{2(k-n-m)}} \frac{\sin^{2n}\theta}{g^{k-m}(\theta)} F_{k,n,m}, \qquad (4.9)$$

with

$$F_{k,n,m} = e^{-k(v_r^2 + v_\theta^2 + v_\phi^2)} v_\phi^{2n} (v_\theta^2 + v_\phi^2)^m. \qquad (4.10)$$

The general solution of the Boltzmann equation can be written as a sum over the basic solutions with arbitrary coefficients. We require the subset of the general solution consistent with the density of Binney's model, which is given by expression (2.6), with $h(\theta)$ given in (2.9). To reproduce the $r$-dependence of this density, we must restrict ourselves to those basic solutions (4.9) with $k = 1 + n + m$. Furthermore, to reproduce the $\theta$-dependence of the density, we must restrict ourselves to just two values of $n$, namely $n = 0$ (i.e., $k - m = 1$), and $n = 1$ (i.e., $k - m = 2$). Implementing these restrictions and integrating over the velocities, we obtain

$$\rho = \frac{1}{r^2} \sum_m \left[ A_{0,m} \frac{1}{g(\theta)} S_{0,0,m} + A_{1,m} \frac{\sin^2\theta}{g^2(\theta)} S_{0,1,m} \right]. \qquad (4.11)$$

Here the $A_{0,m}$ and $A_{1,m}$ are arbitrary amplitudes and the $S_{0,0,m}$ and $S_{0,1,m}$ are the velocity integrals

$$S_{0,n,m} = \iiint F_{1+n+m,n,m} \, dv_r dv_\theta dv_\phi, \qquad (4.12)$$

which can be worked out explicitly (see Appendix B). Finally, to make the density (4.11) implied by the phase-space DF exactly consistent with the density of Binney's model, we have to fulfill two conditions

$$\sum_m A_{0,m} S_{0,0,m} = 2 - Q,$$
$$\sum_m A_{1,m} S_{0,1,m} = 2Q - 2. \qquad (4.13)$$

If we replace the two sets of coefficients $A_{0,m}$ and $A_{1,m}$ by the weights $W_{0,m}$ and $W_{1,m}$ according to

$$A_{0,m} S_{0,0,m} = (2-Q) W_{0,m},$$
$$A_{1,m} S_{0,1,m} = 2(Q-1) W_{1,m}, \qquad (4.14)$$

then the two conditions become

$$\sum_m W_{0,m} = 1, \qquad \sum_m W_{1,m} = 1. \qquad (4.15)$$

The subset of general solutions (4.9) which is consistent with the density of our model can now be written as

$$f = \sum_m \left( A_{0,m} I_1^{m+1} I_3^m + A_{1,m} I_1^{m+2} I_2 I_3^m \right), \qquad (4.16)$$

or, more explicitly, as

$$f = \frac{1}{r^2} \sum_m \left[ \frac{(2-Q)}{g(\theta)} \frac{W_{0,m}}{S_{0,0,m}} e^{-(1+m)(v_r^2 + v_\theta^2 + v_\phi^2)} (v_\theta^2 + v_\phi^2)^m \right.$$
$$\left. + 2 \frac{\sin^2\theta}{g^2(\theta)} (Q-1) \frac{W_{1,m}}{S_{0,1,m}} e^{-(2+m)(v_r^2 + v_\theta^2 + v_\phi^2)} v_\phi^2 (v_\theta^2 + v_\phi^2)^m \right]. \qquad (4.17)$$

This DF should be used with caution. It involves $I_3$ which is a good partial integral for thin and near-thin tubes, but not for fat tubes. The solution (4.16) is exact in the limit in which it contains only terms with $m = 0$ ($W_{0,m} = 0$ and $W_{1,m} = 0$ for $m > 0$). In this limit it depends only on the globally defined exact integrals $I_1$ and $I_2$ given in (4.5) and so reduces to the two-integral DF found earlier by Toomre (1982) and Evans (1993) through another approach.

In the opposite limit of very large $m$, the DF (4.16) also seems trustworthy, for the following reason. For large $m$, the factor $\exp[-m(v_\theta^2 + v_\phi^2)](v_\theta^2 + v_\phi^2)^m$, which occurs in both terms of the DF (4.16), has its maximum at $v_\theta^2 + v_\phi^2 = 1$, and drops off sharply at both sides of this maximum. According to equation (4.8) this defines the part of phase space occupied by thin tubes. Thus, for large $m$, the solution (4.16) concentrates on the phase-space portion in which $I_3$ is an acceptably accurate integral. We conclude that the DF (4.16) may be used safely for any combination of $m = 0$ and large $m$.

### 4.4 Velocity dispersions

The velocity dispersions which correspond to our solution for the phase-space DF can be found as follows. In (4.16), the velocity components occur only in their squares. Accordingly, the off-diagonal terms of the velocity dispersion tensor vanish, i.e.,

$$\langle v_r v_\theta \rangle \equiv 0, \qquad \langle v_r v_\phi \rangle \equiv 0, \qquad \langle v_\theta v_\phi \rangle \equiv 0. \qquad (4.18)$$

The diagonal terms are obtained, as usual, by multiplying $f$ with $v_j^2$, (where $j$ stands for $r, \theta, \phi$, respectively) and integrating over all velocities. We then find

$$\rho \langle v_j^2 \rangle = \frac{1}{r^2} \left[ \frac{2-Q}{g(\theta)} \sum_m W_{0,m} \frac{S_{j,0,m}}{S_{0,0,m}} \right.$$
$$\left. + 2 \frac{\sin^2\theta}{g^2(\theta)} (Q-1) \sum_m W_{1,m} \frac{S_{j,1,m}}{S_{0,1,m}} \right], \qquad (4.19)$$

where we have given the velocity integral (4.12) the more general form

$$S_{j,n,m} = \iiint v_j^2 F_{1+n+m,n,m} \, dv_r dv_\theta dv_\phi. \qquad (4.20)$$

The velocity integrals can be computed analytically by introducing spherical coordinates in the velocity space. This transforms the three-dimensional integral into a product of three one-dimensional integrals (see Appendix B). The results of these computations give for the three velocity dis-

persions

$$\rho \langle v_r^2 \rangle = \frac{1}{r^2} \left[ \frac{2-Q}{g(\theta)} \sum_m \frac{W_{0,m}}{2(1+m)} \right.$$
$$\left. + \frac{\sin^2 \theta}{g^2(\theta)} (Q-1) \sum_m \frac{W_{1,m}}{(2+m)} \right],$$
$$\rho \langle v_\theta^2 \rangle = \frac{1}{r^2} \left[ \frac{2-Q}{2g(\theta)} + \frac{\sin^2 \theta}{2g^2(\theta)} (Q-1) \right], \quad (4.21)$$
$$\rho \langle v_\phi^2 \rangle = \frac{1}{r^2} \left[ \frac{2-Q}{2g(\theta)} + \frac{3\sin^2 \theta}{2g^2(\theta)} (Q-1) \right].$$

Notice that $\langle v_\theta^2 \rangle$ and $\langle v_\phi^2 \rangle$ are independent of the weights $W_{0,m}$ and $W_{1,m}$, and hence are independent of $m$. At the poles, $\langle v_\theta^2 \rangle$ and $\langle v_\phi^2 \rangle$ are equal, while at the equator, $\langle v_\phi^2 \rangle$ is larger than $\langle v_\theta^2 \rangle$. If the weights for $m=0$ are dominant (two-integral DF), the dispersions are isotropic at the poles, while at the equator $\langle v_r^2 \rangle$ and $\langle v_\theta^2 \rangle$ are equal, but $\langle v_\phi^2 \rangle$ is larger. If the weights for large $m$ are dominant (predominance of thin tubes), then $\langle v_r^2 \rangle$ is much smaller than the other two. The weights $W_{0,m}$ and $W_{1,m}$ may be chosen independently of each other so that relatively small radial dispersions can be achieved on the poles or on the equator separately, within limits depending on $Q$. These models are similar to the Type I tangentially anisotropic model 2 of Dehnen & Gerhard (1993). In the strongly flattened limit ($Q=2$, $g(\theta) = 1+\cos^2\theta$), the first term in (4.21) vanishes for each dispersion. In the spherical limit ($Q=1$, $g(\theta)=1$), the second term in (4.21) vanishes for each dispersion and we recover our earlier results (3.6) for the self-consistent isothermal sphere.

## 5 COMPARISON OF JEANS AND BOLTZMANN APPROACHES

### 5.1 Spherically aligned velocity ellipsoid

The comparison of the two-parameter set of solutions (2.28) from the Jeans approach with the results (4.21) from the Boltzmann approach, shows immediately that the terms produced by the second parameter, $H_2$, of the Jeans solution are not reproduced by our approximate Boltzmann solution. This is a direct consequence of the fact that our approximate partial integral contains only the squares of the individual velocity components, but not the cross term $v_r v_\theta$, just as is the case for the two classical integrals. Thus, any distribution function composed of these three integrals must produce a vanishing $\langle v_r v_\theta \rangle$, and spherically aligned velocity ellipsoids. We have already anticipated this circumstance by explicitly writing out the Jeans solutions for the case $H_2 = 0$ in (2.31).

A comparison of equations (4.21) and (2.31) shows that the Jeans approach with $H_2 = 0$ — or $F_2(\theta) \equiv 0$ — and the Boltzmann approach give identical results for $\langle v_\theta^2 \rangle$ and $\langle v_\phi^2 \rangle$, a reassuring circumstance. Now as to $\langle v_r^2 \rangle$, solutions (4.21) and (2.31) will give identical results for all $\theta$, if the coefficients of the $g(\theta)$ terms and, separately, those of the $\sin^2 \theta$ terms are identical. These two conditions can be brought into the form

$$\sum_m \frac{W_{0,m}}{1+m} = 1 + \frac{2H_1}{Q(2-Q)},$$
$$\sum_m \frac{W_{1,m}}{2+m} = \tfrac{1}{2} + \frac{H_1}{Q}. \quad (5.1)$$

There are many ways to satisfy these constraints. For example, let us choose in each sum only two non-vanishing terms, namely one with $m=0$ (two-integral solution) and the other with $m \to \infty$ (thin tube solution). Then

$$W_{0,0} = \alpha_0, \qquad W_{0,\infty} = 1 - \alpha_0,$$
$$W_{1,0} = \alpha_1, \qquad W_{1,\infty} = 1 - \alpha_1, \quad (5.2)$$

which gives for the two sums

$$\sum_m \frac{W_{0,m}}{1+m} = \alpha_0, \qquad 2\sum_m \frac{W_{1,m}}{2+m} = \alpha_1. \quad (5.3)$$

Conditions (5.1) then determine $\alpha_0$ and $\alpha_1$ for any given $H_1$. However $\alpha_0$ and $\alpha_1$ will fall into their physical range (0 to 1) only if

$$-\tfrac{1}{2} Q(2-Q) < H_1 < 0. \quad (5.4)$$

The lower limit for $H_1$ given by condition (5.4) is fulfilled by the Jeans solution according to equation (2.34). However, the upper limit for $H_1$ given by (5.4) is not required by the Jeans solution and arises from our Boltzmann approach, in which we took care of the thin tubes, but not of the fatter ones.

We conclude that the Boltzmann approach with our partial integral reproduces the two-parameter solution of the Jeans equations accurately, but only if $H_2 = 0$, and if $H_1 < 0$. It would seem plausible that these limitations could be raised by, a), developing our third integral to contain a $v_r v_\theta$ term, and by, b), finding another partial third integral that should be of good accuracy for flat tube orbits, with moderate $z$-amplitudes. These two extensions are not attempted here.

### 5.2 The cross-term $\langle v_r v_\theta \rangle$

The general solution (2.17) of the Jeans equations for oblate scale-free potentials contains many solutions with $F_2(\theta) \neq 0$, i.e., $\langle v_r v_\theta \rangle \neq 0$, which have velocity ellipsoids that are not aligned with the spherical coordinates $r$ and $\theta$. Such a term arises when the DF depends on a third integral which contains a term proportional to $v_r v_\theta$. Might a more accurate partial third integral than our $I_3 = L^2 g(\theta)$ contain such a term? We have not found the overall answer to this question. The early numerical experiments by Richstone (1984) for the oblate scale-free logarithmic potential suggest that many self-consistent solutions show misalignment of the velocity ellipsoid relative to $(r, \theta)$. Levison & Richstone (1985a, b) found the same for $r^{-3}$ densities in Binney's potential. Here we add a simple argument indicating how such a misalignment might come about.

Consider first the solution constructed exclusively from thin tubes. Such solutions have been derived in detail for Stäckel potentials (Bishop 1987; de Zeeuw & Hunter 1990). A thin tube has a meridional cross section which is elongated — larger $r$ — in the polar direction of the oblate potential (see orbits A, B, C, and D in Fig. 3). The velocity vector $(v_r, v_\theta)$ must fall along this cross section. The polar elongation then gives $v_r \cdot v_\theta < 0$ for all points above the equator (the opposite sign holds for points below the equator). Thus for the thin-tube solution $\langle v_r v_\theta \rangle < 0$ above the equator. Note that this sign for $\langle v_r v_\theta \rangle$ holds for the scale-free logarithmic potential and is therefore not caused by the existence of a regular core, such as occurs in Stäckel models.

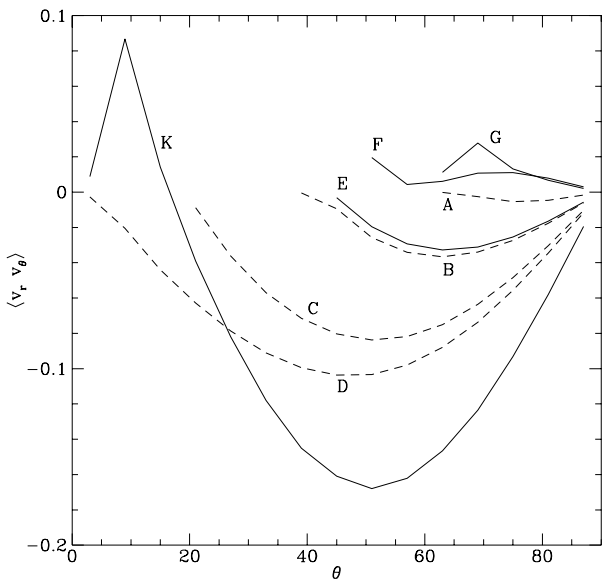

**Figure 5.** Contributions to $\langle v_r v_\theta \rangle$ by individual orbits as a function of the angle $\theta$. The near-thin orbits A, B, C, D all contribute negative values for all $\theta$. They are shown in a broken line. Very fat orbits, such as F, G and K, contribute both positive and negative values. They are shown in an unbroken line.

In contrast, the two-integral solution, which contains all orbits, not just the thin tubes, has $\langle v_r v_\theta \rangle \equiv 0$. Do the thicker tubes then contribute positively to $\langle v_r v_\theta \rangle$ above the equator? The answer seems to be yes, as indicated by the following evidence.

A solution for a DF in a given potential can be thought of as consisting of the sum of contributions from all the individual orbits involved. In an oblate potential a tube orbit provides a $\langle v_r v_\theta \rangle$ contribution that is a function of $r$ and $\theta$. In a scale-free oblate potential, for a scale-free DF, a specific orbit can be taken to represent a sequence of similar orbits, all equal except in scale, i.e., in energy (Richstone 1980; Schwarzschild 1993). This whole sequence is taken into account by projecting the integrated orbit onto a reference circle in the meridional plane, with the appropriate scaling rules. For the logarithmic scale-free potential the rules are: Leave the velocities unchanged and weigh each point in the integrated orbit by $dt/r$. Thus the resulting $\langle v_r v_\theta \rangle$ for each orbit, i.e., orbit sequence, depends only on $\theta$.

Results of $\langle v_r v_\theta \rangle$ computations are given in Figure 5 for the seven orbits of Table 2 with $q = 0.9$. The four thin tubes A, B, C, and D give negative contributions to $\langle v_r v_\theta \rangle$ for the entire range of $\theta$, just as expected from the geometrical argument given above. The absolute values of $\langle v_r v_\theta \rangle$ for these thin tubes are relatively small, generally less than 10% of $\langle v_\theta^2 \rangle$, leading to less than $6°$ of tilt for the velocity ellipsoid in thin-tube solutions. For lower $q$ values (galaxies flatter than E3) the maxima of $|\langle v_r v_\theta \rangle|$ are higher, reaching about 20% for $q = 0.8$, and still larger values for smaller $q$, such as used by Richstone (1984) and Levison & Richstone (1985a).

Now to the thicker tubes E, F, and G. Orbit E, the thinnest of the three, is still thin enough to give all negative $\langle v_r v_\theta \rangle$, like the thin tubes. Orbit F shows in Figure 5 the first positive $\langle v_r v_\theta \rangle$, but still very small. Only orbit G of all shown in Figure 3 makes substantial positive contributions to $\langle v_r v_\theta \rangle$, and that mainly near its upper margin (near the smallest $\theta$ it reaches). To show that orbit G is not a rare case, we have added a very thick orbit K (its upper turning point is plotted in Figure 3). Figure 5 shows that orbit K makes strong positive contributions to $\langle v_r v_\theta \rangle$, again concentrated near its upper margin. This shows explicitly that thick orbits like G and K balance the thin orbits to give $\langle v_r v_\theta \rangle = 0$ for the two-integral solutions.

Can we understand the positive contributions of the thick orbits in simple geometrical terms? The direction of the upper margins of tube orbits in the meridional cross section shown in Figure 3 gives the answer. These upper margins are not exaclty radially directed but are slightly turned clockwise — reminiscent of the ellipsoidal coordinates that define the margins of orbits in a Stäckel potential. This tilt in direction gives a positive $\langle v_r v_\theta \rangle$ on the upper margin. In the neighbourhood of this margin the velocity vector of a thick tube orbit straddles the direction of the margin. Hence near the upper margin a thick tube orbit should contribute a positive $\langle v_r v_\theta \rangle$, just as Figure 5 shows for orbits G and K.

We conclude that for the oblate scale-free logarithmic potential: A solution made up of thin or near-thin tubes — tangentially anisotropic — will have a negative $\langle v_r v_\theta \rangle$ above the equatorial plane, though of modest absolute amount. In contrast, a solution made up of thick tubes — radially anisotropic — may have positive $\langle v_r v_\theta \rangle$ above the equator, at least in part of the meridional plane.

## 6 OBSERVABLES

### 6.1 General scale-free case

We define a coordinate system $(x', y', z')$ with $z'$ along the line of sight by the transformation

$$\begin{pmatrix} x' \\ y' \\ z' \end{pmatrix} = \begin{pmatrix} 0 & 1 & 0 \\ -\cos i & 0 & \sin i \\ \sin i & 0 & \cos i \end{pmatrix} \begin{pmatrix} x \\ y \\ z \end{pmatrix}. \quad (6.1)$$

Here $i$ is the inclination, defined such that $i = \pi/2$ corresponds to edge-on observation. Integration of model properties over $z'$ then gives the projected (observable) properties on the sky-plane $(x', y')$. General expressions for the line-of-sight projected properties can be found in Appendix A of Evans & de Zeeuw (1994).

The projected properties of scale-free models with potentials (2.5) and densities (2.6) are themselves scale-free. It is therefore useful to define polar coordinates $(R', \Theta')$ in the plane of the sky by

$$x' = -R' \sin \Theta', \qquad y' = R' \cos \Theta', \quad (6.2)$$

so that $R'^2 = x'^2 + y'^2$ and $\theta'$ runs counterclockwise, and is measured from the $y'$-axis ("North"). Then the projected surface density is

$$\Sigma(R', \Theta') = \frac{H(\Theta')}{(R')^{\gamma-1}}, \quad (6.3)$$

**Table 3.** Non-vanishing coefficients in expression (6.7) for the projected second moments.

| Coefficient | General | Edge-on |
|---|---|---|
| $a_{11}$ | $3 - q'^2 q^{-2} + 2H_1 q'^2$ | $2(1 + H_1 q^2)$ |
| $b_{11}$ | $1 + q^{-2}$ | $1 + q^{-2}$ |
| $a_{22}$ | $1 - q'^2 q^{-2} + 2q'^4 q^{-2}$ | $2q^2$ |
| $b_{22}$ | $3 - q^{-2} + 2H_1 q'^2$ | $3 - q^{-2} + 2H_1 q^2$ |
| $a_{33}$ | $1 + 2H_1 q^2 + q'^2 q^{-2}(1 + 2q^2 - 2q'^2)$ | $2 + 2H_1 q^2$ |
| $b_{33}$ | $3 - q^{-2} + 2H_1(q^2 + 1 - q'^2)$ | $3 - q^{-2} + 2H_1$ |
| $a_{23}$ | $-2q'^2 q^{-2}(1 - q'^2)^{1/2}(q'^2 - q^2)^{1/2}$ | $0$ |
| $b_{23}$ | $-2H_1(1 - q'^2)^{1/2}(q'^2 - q^2)^{1/2}$ | $0$ |
| $c_{12}$ | $2(1 - q'^2 q^{-2} - H_1 q'^2)$ | $-2H_1 q^2$ |
| $c_{13}$ | $2(q^{-2} - H_1)(1 - q'^2)^{1/2}(q'^2 - q^2)^{1/2}$ | $0$ |

with

$$H(\Theta') = \int_{-\infty}^{\infty} \frac{ds}{s^\gamma} \frac{h(\theta[\Theta', s])}{g^2(\theta[\Theta', s])}, \qquad (6.4)$$

where

$$\tan^2 \theta = \frac{(s \sin i - \cos \Theta' \cos i)^2 + \sin^2 \Theta'}{(s \cos i + \cos \Theta' \sin i)^2}. \qquad (6.5)$$

Similar expressions for the projected velocity moments can be written down using equations in Appendix A of Evans & de Zeeuw (1994).

### 6.2 Binney's model

The motivation for introducing the two-parameter Jeans solution (2.28) for Binney's model in Section 2.6 is that the line-of-sight integrations can be done exactly to give the observables. We have shown that a one-parameter subset of these Jeans solutions ($H_2 = 0$ and $H_1 < 0$) corresponds to physical DFs, at least to good approximation. Let us now see how these galaxy models appear to an observer.

The projected surface density of Binney's model is

$$\Sigma(x', y') = \frac{\pi q}{q'^3} \frac{x'^2 + y'^2}{[x'^2 + y'^2 q'^{-2}]^{3/2}}$$
$$= \frac{\pi q}{q'^3 R'} \frac{1}{(\cos^2 \Theta' + q'^{-2} \sin^2 \Theta')^{3/2}}, \qquad (6.6)$$

with $q'^2 = \cos^2 i + q^2 \sin^2 i$. The six components of the projected second moments follow from equation (A7) of Evans & de Zeeuw (1994). The line-of-sight second moment $\langle v_{z'}^2 \rangle = \langle v_{\rm los}^2 \rangle = \langle \mu_{z'z'}^2 \rangle$ is the best known of these quantities. The other two diagonal components $\langle \mu_{x'x'}^2 \rangle$ and $\langle \mu_{y'y'}^2 \rangle$ are the dispersions in the proper motions. The off-diagonal components, $\langle \mu_{x'y'}^2 \rangle$, $\langle \mu_{y'z'}^2 \rangle$ and $\langle \mu_{x'z'}^2 \rangle$ measure the correlations between the distributions of proper and radial velocities. All these quantities are of the form

$$\langle \mu_{x_i x_j}^2 \rangle = \frac{a_{ij} x'^2 + b_{ij} y'^2 + c_{ij} x' y'}{4(x'^2 + y'^2)}, \qquad (6.7)$$
$$= \tfrac{1}{4}(a_{ij} \cos^2 \Theta' + b_{ij} \sin^2 \Theta' + c_{ij} \cos \Theta' \sin \Theta'),$$

where the non-vanishing $a_{ij}$, $b_{ij}$, and $c_{ij}$ depend on $i$, $q$ and $H_1$ and are given in Table 3.

The velocity profile (VP) is the distribution of line-of-sight velocities $v_{\rm los} = v_{z'}$ measured at position $(x', y')$ on the sky (see, e.g., van der Marel & Franx 1993; Gerhard 1993). It is obtained by integrating the DF along the line-of-sight and over the tangential velocity components (see e.g., Evans & de Zeeuw 1994). The normalised VP is

$$\mathrm{VP}(v_{z'}, x', y') = \frac{1}{\Sigma(x', y')} \int_{-\infty}^{\infty} dz' \int_{-\infty}^{\infty} dv_{x'} \int_{-\infty}^{\infty} dv_{y'} f(I_1, I_2, I_3). (6.8)$$

Recently, there have been a number of investigations of the range of shapes of the VPs of flattened galaxy models with two-integral DFs (e.g., Evans 1994; Qian et al. 1995; Carollo et al. 1995). The variety of shapes of the VPs of three-integral DFs is expected to be richer still (e.g., Dehnen & Gerhard 1993).

First, we consider the VPs of the DF (4.16). When only the $m = 0$ terms are present, it reduces to the two-integral DF. In the non-rotating case, the VP of the two-integral DF is the sum of two Gaussians, as shown in Appendix B of Evans & de Zeeuw (1994). It is useful to contrast this with the VP of a model dominated by near-thin tube orbits, say with just the $m = 20$ terms in (4.16). Fig. 6a shows the VPs observed on the major axis for edge-on viewing ($\theta = 90°$) in these two cases. The intrinsic flattening of the model is E3 ($q = 0.9$). When this galaxy is built from near-thin tubes, the VP becomes squatter. The wings of the VP become shallower as there are fewer eccentric orbits, while there is a sudden and substantial rise in the VP at the circular speed. In the commonly-used Gauss-Hermite decomposition (see e.g., van der Marel & Franx 1993), the VPs of the near-thin tube models possess negative values for the $h_4$ coefficient. The dotted curve shows the VP of the $m = 1$ component DF, which has properties intermediate between the two-integral and near-thin tube DFs.

Second, we investigate the VPs of the DFs corresponding to our Jeans solutions. We choose the simplest conditions (5.2). For any flattening, there is a one-parameter family of Jeans solutions depending on $H_1$. For any Jeans solution, there is a one-parameter sequence of DFs, depend-ing on $m$. In other word, the DF includes a contribution from the two-integral DF ($m = 0$ component), together with a contribution from a component at one other $m$. Fig. 6b shows the VPs generated when the second component is $m = 1$ (full curve) and $m = 20$ (dashed curve). Both VPs have the same second moments and correspond to the same Jeans solution (2.31), which is tangentially anisotropic ($H_1 = -0.45$). The effect of including lots of thin and near-thin tubes in the DFs is to produce characteristic kinks in the VP at roughly the circular speed. An observer sees substantial numbers of thin and near-thin tubes with $v_\phi \approx \pm 1$ and with $v_\theta \approx \pm 1$. When the models are viewed edge-on, this produces kinks in the VP at $v_{z'} \approx \pm 1$ as well as the peak at $v_{z'} \approx 0$. When the models are viewed pole-on, the observer sees mainly the $v_\theta$ velocity components of orbits at the tangent point. So, the characteristic appearance of the VP of a model dominated by near-thin tubes is robust under inclination angle changes.

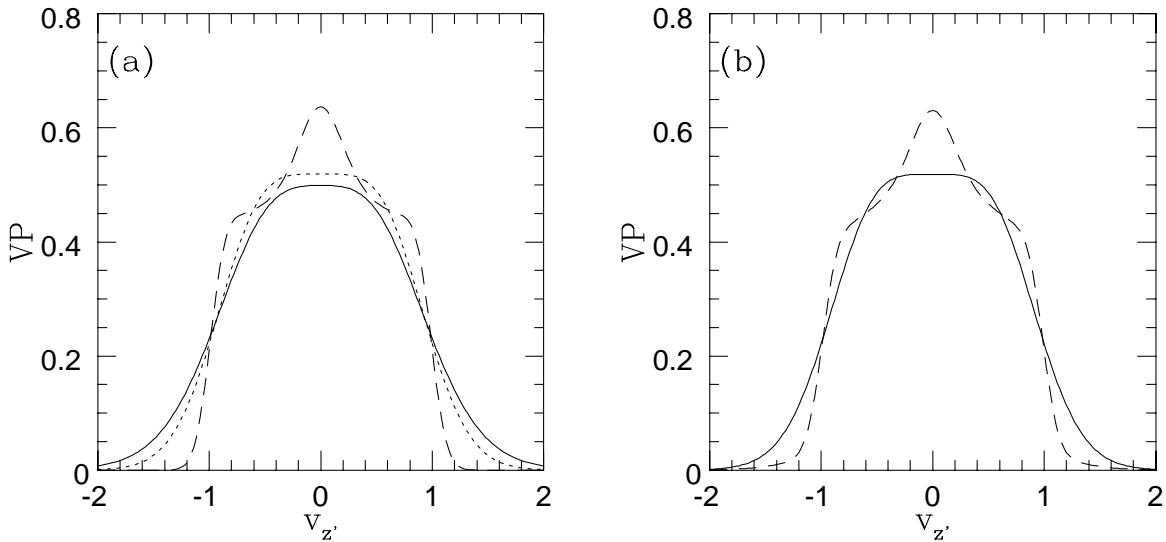

**Figure 6.** (a) VPs of Binney's model with an ellipticity of E3 viewed edge-on ($i = 90°$). The viewing point is on the projected major axis. The full curve is the VP generated by the two-integral DF ($m = 0$ in (4.16)). The dashed curve is the VP of a model dominated by near-thin tubes ($m = 20$). The dotted curve shows an intermediate case ($m = 1$). (b) VPs of Binney's model which are consistent with the Jeans solutions (2.31) with $H_1 = -0.45$. The full curve is built from the $m = 0$ and $m = 1$ components, the dashed curve from the $m = 0$ and $m = 20$ components.

All the VPs of the models with many near-thin tubes have negative $h_4$ coefficients in the Gauss-Hermite decomposition. This result is valid for a wider range of flattened scale-free models. For example, Carollo et al. (1995) showed that the VPs of $f(E, L_z)$ non-rotating scale-free spheroids with a range of density profile slopes have negative $h_4$ coefficients. Increasing the tangential anisotropy by populating more near-thin tube orbits at the expense of the fat tubes will lead to $h_4$-values that are more negative.

## 7 CONCLUSIONS

We have derived the general solution of the Jeans equations for arbitrary oblate scale-free densities in oblate scale-free logarithmic potentials. This provides all possible second moments $\langle v_r^2 \rangle$, $\langle v_r v_\theta \rangle$, $\langle v_\theta^2 \rangle$ and $\langle v_\phi^2 \rangle$ that can support the given density in the gravitational field generated by the potential. The general solution contains two free functions, $F_1(\theta)$ and $F_2(\theta)$, which give the angular variation of $\langle v_r^2 \rangle$ and $\langle v_r v_\theta \rangle$, respectively. We have studied the specific case of Binney's model in detail. This is the self-consistent scale-free model with spheroidal equipotentials that is often used as a simple dark halo model. We have identified a simple two-parameter set of Jeans solutions for this model with the useful property that the observable kinematic quantities can be given explicitly.

In order to identify which of these Jeans solutions correspond to physical phase-space distribution functions, we first investigated the spherical limit. In this case the potential reduces to that of the singular isothermal sphere, and all physical Jeans solutions for scale-free densities of arbitrary flattening must have $F_2(\theta) \equiv 0$, i.e., $\langle v_r v_\theta \rangle \equiv 0$, so that the velocity ellipsoids must be spherically aligned. We have provided DFs for all the remaining Jeans solutions.

We then investigated the problem of finding distribution functions associated with the Jeans solutions in the flattened scale-free logarithmic potentials of Binney's model. We found an approximate solution of the collisionless Boltzmann equation, and showed by numerical orbit integration that it provides a third (partial) integral of good accuracy for thin and near-thin tube orbits. It is a modification of the total angular momentum. We then constructed simple three-integral distribution functions, and showed that the kinematic properties of these approximate DFs agree with the subset of our two-parameter Jeans solutions which have spherically aligned velocity ellipsoids (i.e., the parameter $H_2(\theta) \equiv 0$). Although our DFs are approximate, this result suggests that the properties of the moderately flattened anisotropic scale-free Binney models can be approximated quite accurately by assuming spherical alignment of the velocity ellipsoid.

The derivation presented in Section 2 suggests that the general solution of the higher-order moment equations may be found in a similar way. We expect that at each order further free functions of $\theta$ appear. The problem of picking those that correspond to physical DFs remains. One might also attempt to solve the entire collisionless Boltzmann equation directly. White (1985) was able to find the general solution for scale-free spherical models because in this geometry three exact integrals of motion are known. Since there is no exact third integral for an arbitrary axisymmetric scale-free potential, an exact solution of the collisionless Boltzmann equation seems to be precluded, even though it may be possible to find solutions of the moment equations to arbitrary high order. It is still possible that exact, globally defined third integrals may exist for some specific choices of $g(\theta)$.

Many of the properties of the two-integral Binney model carry over to the general family of axisymmetric power-law galaxies (Evans 1994), which have potentials of the form

$$\Phi = \begin{cases} \frac{1}{2}\ln(R_c^2 + R^2 + z^2/q^2), & \text{for } \beta = 0, \\ -\frac{R_c^\beta}{\beta(R_c^2 + R^2 + z^2/q^2)^{\beta/2}}, & \text{for } \beta \neq 0. \end{cases} \quad (7.1)$$

Compared with the case considered in this paper, these models have two extra parameters: the power $\beta$ of the radial dependence of the spheroidal potential, and the core radius $R_c$. The general solution of the Jeans equations given here for Binney's model ($R_c = 0, \beta = 0$) can be generalized to scale-free cases with $\beta \neq 0$. When $R_c \neq 0$ the radial dependence of the stresses is not known a priori, and solving the Jeans equations becomes a harder problem. Our results suggest that assuming spherical alignment of the velocity ellipsoid might be a good approximation over most of the model, and the general solution with $\langle v_r v_\theta \rangle \equiv 0$ is available (Bacon 1985). The two-parameter set of solutions presented in Section 2.6 can be generalized to the $\beta \neq 0, R_c \neq 0$ case, and the observable properties of these solutions can again be written down explicitly. The partial integral presented here for Binney's model can also be modified to the general power-law models. These results will be discussed in a future paper.

A further generalization is to consider triaxial models. The axisymmetric power-law galaxies can be made triaxial by taking the equipotential surfaces to be similar triaxial ellipsoids rather than spheroids. There are then three Jeans equations for the six components of the stress tensor. The scale-free case can be treated analogously as we have done here, and the two-parameter solution can be generalized as well. We expect it to contain three free parameters. Analysis of this interesting case is in progress.

## ACKNOWLEDGMENTS

It is a pleasure to thank Marcella Carollo for stimulating discussions and for comments on the manuscript. TdZ acknowledges the hospitality of the Institute for Advanced Study, where part of this work was done with the support of NSF grant PHY 92–45317. NWE thanks the Leids Kerkhoven Bosscha Fonds for a travel grant which made a visit to Leiden possible. He is supported by the Royal Society.

## APPENDIX A: GENERAL SOLUTION IN CYLINDRICAL COORDINATES

In many studies, properties of axisymmetric dynamical models are given in terms of standard cylindrical coordinates $(R, \phi, z)$. Our general solution (2.17) results in the following components of the stress tensor in cylindrical coordinates:

$$\begin{aligned} \rho\langle v_R^2 \rangle &= \frac{G_1(\theta)}{r^\gamma g^2(\theta)}, & \rho\langle v_R v_z \rangle &= \frac{G_2(\theta)}{r^\gamma g^2(\theta)}, \\ \rho\langle v_z^2 \rangle &= \frac{G_3(\theta)}{r^\gamma g^2(\theta)}, & \rho\langle v_\phi^2 \rangle &= \frac{G_4(\theta)}{r^\gamma g^2(\theta)}, \end{aligned} \quad (A1)$$

$$G_1(\theta) = F_1(\theta)\sin^2\theta + 2F_2(\theta)\sin\theta\cos\theta + F_3(\theta)\cos^2\theta,$$
$$G_2(\theta) = [F_1(\theta) - F_3(\theta)]\sin\theta\cos\theta$$
$$+ F_2(\theta)[\cos^2\theta - \sin^2\theta], \quad \text{(A2)}$$
$$G_3(\theta) = F_1(\theta)\cos^2\theta - 2F_2(\theta)\sin\theta\cos\theta + F_3(\theta)\sin^2\theta.$$

Spherically aligned solutions have $F_2(\theta) \equiv 0$. Cylindrically aligned solutions, $\langle v_R v_Z \rangle \equiv 0$, must have $G_2(\theta) \equiv 0$, and hence are obtained by taking $F_1(\theta) = I(\theta) + F_2(\theta)\tan\theta$, with $F_2(\theta)$ arbitrary. Then

$$G_1(\theta) = I(\theta) + \frac{F_2(\theta)}{\sin\theta\cos\theta}, \qquad G_3(\theta) = I(\theta). \quad \text{(A3)}$$

Taking also $F_2(\theta) \equiv 0$ gives the $f = f(E, L_z)$ solution with $\langle v_R^2 \rangle \equiv \langle v_z^2 \rangle$ and $\langle v_R v_z \rangle \equiv 0$.

## APPENDIX B: AN AUXILIARY INTEGRAL

The normalising constant in our three-integral DF involves the integrals (4.12):

$$S_{0,n,m} = \iiint dv_r \, dv_\theta \, dv_\phi$$
$$e^{-(1+n+m)(v_r^2 + v_\theta^2 + v_\phi^2)} v_\phi^{2n} (v_\theta^2 + v_\phi^2)^m. \quad \text{(B1)}$$

To evaluate, we introduce spherical polars in velocity space (see BT, p. 293)

$$v_r = v\cos\eta, \quad v_\theta = v\sin\eta\cos\psi, \quad v_\phi = v\sin\eta\sin\psi. \quad \text{(B2)}$$

This reduces the three-dimensional integral to a product of three one-dimensional integrals, which can be evaluated using formulae 3.621.5 and 3.381.4 of Gradshteyn & Ryzhik (1980). We finally obtain:

$$S_{0,n,m} = \frac{\pi\Gamma(n+\tfrac{1}{2})\Gamma(n+m+1)}{(m+n+1)^{m+n+\tfrac{3}{2}}\Gamma(n+1)}. \quad \text{(B3)}$$

The generalised velocity integrals (4.20) can be found in a similar manner. We find

$$S_{1,m,n} = \frac{S_{0,m,n}}{2(m+n+1)},$$
$$S_{2,m,n} = \frac{S_{0,m,n}}{2(n+1)}, \quad \text{(B4)}$$
$$S_{3,m,n} = \frac{(2n+1)S_{0,m,n}}{2(n+1)}.$$

This paper has been produced using the Blackwell Scientific Publications TEX macros.